\documentclass[conference]{IEEEtran}
\IEEEoverridecommandlockouts

\usepackage{cite}
\usepackage{hyperref}
\usepackage{amsmath,amssymb,amsfonts}
\usepackage{algorithmic}
\usepackage{graphicx}
\usepackage{textcomp}
\usepackage{xcolor}
\usepackage{gensymb}
\usepackage{subcaption}
\usepackage{caption}

\def\BibTeX{{\rm B\kern-.05em{\sc i\kern-.025em b}\kern-.08em
    T\kern-.1667em\lower.7ex\hbox{E}\kern-.125emX}}
\begin{document}

\title{Emphasis Sensitivity in Speech Representations
\thanks{This work was supported by the UKRI AI Centre for Doctoral Training in Speech and Language Technologies (SLT) and their Applications, funded by UK Research and Innovation [grant number EP/S023062/1], with additional support from Huawei Research \& Development (UK).
We thank Nicola Mendini and Mattias Cross for their proofreading and helpful discussions.}
}

\author{Shaun Rafael Cassini, Thomas Hain, Anton Ragni\\
University of Sheffield, Sheffield, UK\\
\small \texttt{\{srcassini1, t.hain, a.ragni\}@sheffield.ac.uk}
}

\maketitle

\IEEEpubid{\begin{minipage}{\textwidth}\vspace{80pt}
\centering \scriptsize
For the purpose of open access, the author has applied a Creative Commons Attribution (CC BY) license to any Author Accepted Manuscript version arising.\\
© 2025 IEEE. Accepted to IEEE ASRU 2025. Final version at IEEE Xplore: [DOI TBA.]
\end{minipage}}

\begin{abstract}

This work investigates whether modern speech models are sensitive to prosodic emphasis---whether they encode emphasized and neutral words in systematically different ways.
Prior work typically relies on isolated acoustic correlates (e.g., pitch, duration) or label prediction, both of whichwhich miss the relational structure of emphasis.
This paper proposes a residual-based framework, defining emphasis as the difference between paired neutral and emphasized word representations. 
Analysis on self-supervised speech models shows that these residuals correlate strongly with duration changes and perform poorly at word identity prediction, indicating a structured, relational encoding of prosodic emphasis.
In ASR fine-tuned models, residuals occupy a subspace up to 50\% more compact than in pre-trained models, further suggesting that emphasis is encoded as a consistent, low-dimensional transformation that becomes more structured with task-specific learning.

\end{abstract}

\vspace{-1em}

\section{Introduction}

Speech conveys much more than words, as it carries information about the speaker, their mood, and communicative intent.
In particular, speakers use emphasis to highlight specific words or phrases, conveyed through a combination of prosodic cues such as pitch, duration, and loudness.
Prior studies show that prosody in speech enables a listener to recover cues that signal the communicative function of an utterance \cite{herment_pragmatic_2012}. 
Emphasis serves a range of communicative functions, including marking contrast, highlighting information structure, and resolving syntactic ambiguity which words alone may not express \cite{pierrehumbert_meaning_1990, lauerbach_emphasis_2011, wagner_prosodic_2020}.
Automatic speech processing systems that are sensitive to emphasis cues are known to perform better on tasks ranging from intent prediction \cite{rajaa_improving_2023}, speech translation \cite{tsiamas_speech_2024}, to text-to-speech synthesis (TTS) \cite{joly_controllable_2023}. 
Yet it remains unclear to what extent emphasis is implicitly learned by such systems.

Emphasis is expressed in several prosodic cues (acoustic correlates) and is suprasegmental, spanning multiple speech segments in an utterance \cite{van_heuven_phonetic_2021, kohler_what_2006}. 
Its realization is known to vary by speaker, utterance, language, and dialect \cite{cole_new_2016, ladd_prosodic_2023}. 
Cue-specific models which extract acoustic correlates such as the fundamental frequency $\mathrm{F}_0$ can perform well when emphasis aligns with that cue \cite{arons_pitch-based_1994}. 
However, they may miss instances where emphasis is conveyed through under-modeled cues \cite{kochanski_loudness_2005}. 
For instance, post-focal compression refers to a reduction in prosodic cues on segments following an emphasized word.
This representation of emphasis requires modeling context that extends beyond the emphasized word itself \cite{xu_prosodic_2012}. 
Approaches based on local acoustic correlates would be blind to such cues.

Given the limitations of cue-specific approaches and the distributed nature of emphasis, recent work focuses on supervised models with trainable parameters that map the speech signal to emphasis labels.
Some approaches make direct use of the waveform \cite{de_seyssel_emphassess_2024, vaidya_deep_2022}, while others jointly learn acoustic correlates like $\mathrm{F}_0$ or spectral energy \cite{zhang_emphasis_2018, shechtman_supervised_2021}. 
Such supervised approaches require emphasis labels, yet the occurrence of emphasis in natural speech is scarce and labeling data is highly subjective---shaped by the same perceptual ambiguities they aim to model \cite{roettger_mapping_2019}.
This motivates the question: \textit{To what extent do modern speech models, trained without supervision for emphasis, implicitly encode it?}

Prior work has examined acoustic correlates of emphasis or trained classifier probes to predict emphasis labels on individual words \cite{yang_what_2023, de_seyssel_emphassess_2024}.
Both strategies ignore the fact that emphasis is inherently relational: a word sounds prominent only relative to how it would sound without emphasis (neutral) and to the prosodic context around it \cite{turnbull_prominence_2017}.
This work therefore probes for emphasis sensitivity in the residual space between representations of paired neutral-and-emphasized words.
An analysis is conducted to assess whether the residual space encodes emphasis as a consistent, learnable relationship rather than a property of isolated words.

In this work, a residual space is derived from representations extracted from multiple self-supervised speech learning models (S3L models) \cite{mohamed_self_2022} and their fine-tuned variants, specifically those fine-tuned for automatic speech recognition (ASR) and emphasis classification.
Both model types capture prosodic cues in their representations \cite{vitale_exploring_2024, lin_utility_2023}, making them strong candidates for investigating whether emphasis sensitivity arises implicitly across distinct objectives.
This also enables a direct comparison between S3L models and their fine-tuned counterparts to assess how training objectives shape their emphasis sensitivity.
Experiments are conducted on 3,732 word pairs derived from a synthetic dataset designed for emphasis control \cite{de_seyssel_emphassess_2024}, comprising contrastive pairs of utterances with emphasized and neutral words. 
Representations are extracted from multiple state-of-the-art S3L models and their fine-tuned variants. 
Through analysis of residual vectors between representations derived from neutral--emphasized word pairs, this work finds that emphasis is encoded as a low-dimensional, consistent transformation that becomes more pronounced in fine-tuned models.
The contributions of this work are as follows:

\begin{itemize}
    \item A novel framework for quantifying emphasis sensitivity in speech models as a structured relationship between emphasized and neutral word pairs.
    \item A residual-based probing analysis to isolate and measure prosodic variation in representation space.
    \item Experiments showing that S3L models exhibit structured, layer-dependent emphasis sensitivity, which becomes stronger and more consistent after ASR fine-tuning.
    \item Evidence that residuals capture the shift from neutral to emphasized words, occupy a significantly lower-dimensional subspace, and correlate well with duration change.
    \item A geometric interpretation of how emphasis is encoded across architectures, offering insights for emphasis-aware model development.
\end{itemize}

\section{Related Work}
\label{sec:background}

In practice, sensitivity to emphasis has been shown to benefit a variety of downstream tasks, including intent prediction \cite{rajaa_improving_2023}, emotion recognition \cite{kammoun_pitch_2006, cao_prosodic_2014}, ASR \cite{chen_prosody_2006}, naturalistic transcription \cite{cho_leveraging_2022, tilk_lstm_2015}, voice conversion \cite{nguyen_high_2016}, speech segmentation \cite{shriberg_prosody_2000, arons_pitch-based_1994}, speech translation \cite{do_improiving_2015, do_preserving_2017, tsiamas_speech_2024, tsiartas_study_2013}, human-machine dialogue \cite{marge_spoken_2022, velner_intonation_2020, beier_marking_2024}, TTS \cite{strom_modelling_2007, latif_controlling_2021, joly_controllable_2023}, and assisting with language learning \cite{levis_teaching_2004, matzinger_influence_2021, banno_proficiency_2023}. 
For instance, modeling pitch-based emphasis cues reduced the word error rate of an HMM-based ASR system on the Boston News Corpus by 11\% relative to prosody-independent systems \cite{chen_prosody_2006}.

Explicit approaches to emphasis detection rely on supervised learning or acoustic cue extraction. 
These include classifiers trained on prosodic correlates such as pitch, duration, and energy \cite{zhang_emphasis_2018, shechtman_supervised_2021} or on direct waveform inputs \cite{vaidya_deep_2022, de_seyssel_emphassess_2024}. 
Some also incorporate explicit theoretically grounded feature engineering, e.g., $\mathrm{F}_0$ contours or post-focal compression effects \cite{arons_pitch-based_1994, kohler_what_2006}.

\subsubsection{Representation Space Analysis}
Recent studies have shown that S3L models, such as wav2vec 2.0 \cite{baevski_wav2vec2_2020}, HuBERT \cite{hsu_hubert_2021}, and WavLM \cite{chen_wavlm_2022} encode information about prosodic structure, which includes prominence and intonation \cite{vitale_exploring_2024, lin_utility_2023, yang_superb_2021}.
However, such findings often rely on supervised classifier probes trained to map intermediate representations to prosodic labels \cite{yang_what_2023, de_la_fuente_layer-wise_2024}. 
Such probes introduce learnable parameters that risk overstating or misattributing what is encoded versus what is merely decodable \cite{belinkov_probing_2022}. 

\subsubsection{Analysis of Residual Spaces}

Since this work probes for emphasis sensitivity in residual representation space, it is useful to consider how residual analysis has been applied in related domains.
In anomaly detection, PCA-based residual analysis is used under the assumption that structured data lie in a low-dimensional subspace, while residuals represent deviations or noise\cite{kanda_admire_2013}. 
PCA separates signal components that are compressible from those that are distributed or unstructured. 
In contrast, this work does not treat residuals as anomalies, but instead asks whether the residuals themselves reflect a consistent transformation by exhibiting low-rank structure.

A related framework is residual component analysis (RCA) \cite{kalaitzis_residual_2012}, which models structure in the residual covariance after accounting for known variation. 
RCA has been used to recover latent dynamics such as skeletal motion from residuals in motion capture data. 
While this work does not adopt RCA’s probabilistic formulation, it shares the underlying view that residuals can encode meaningful, interpretable structure;
the perspective applied to prosodic emphasis in speech in this proposed analysis.

\section{Experimental Setup \& Methods}
\label{sec:method}

\subsection{Data}
\subsubsection{Synthetic Emphasis Dataset}
The dataset used in this work is derived from the EmphAssess evaluation dataset, a benchmark for evaluating emphasis preservation in speech-to-speech models \cite{de_seyssel_emphassess_2024}.
EmphAssess is comprised of variants of 299 sentences, with each variant changing which word is emphasized, as shown by the example in Figure~\ref{fig:dataset-snippet}.
This yields 913 unique sentence variations.
The sentences are synthesized with four American TTS voices (2 male, 2 female), yielding 3,652 short utterances (2.42 hours).

\begin{figure}[!h]
    \centering
    \small
    \begin{enumerate}
        \centering
        \item \textit{``\underline{\textbf{The}} dishonest \underline{politician} who \underline{admits} it?"}
        \item \textit{``\underline{The} dishonest \underline{\textbf{politician}} who \underline{admits} it?"}
        \item \textit{``\underline{The} dishonest \underline{politician} who \underline{\textbf{admits}} it?"}
    \end{enumerate}
    \caption{Example sentence from the EmphAssess dataset, with neutral words underlined and emphasized words in bold.}
    \label{fig:dataset-snippet}
\end{figure}

There are 546 unique neutral--emphasized word pairs.
Of the 13,108 total words instances across all speakers and transcripts, 3,796 (0.52 hours) are emphasized and 9,312 (0.80 hours) are neutral.
Analysis is conducted on representations of these words, as explained in the following section.

\subsubsection{Deriving Word-Level Representations}
\label{sec:representation-extraction}
As this work examines word--word comparisons, the following describes the method used to obtain word-level representations. 
First, model outputs are aligned to time-stamped word boundaries, following the procedure described in \cite{pasad_what_2024}.
Word-level time boundaries are obtained using the Montreal Forced Aligner (MFA) \cite{mcauliffe_montreal_2017}.  
All frames associated with a specific word are averaged to obtain a representation at each encoder layer, denoted by $\mathbf{z}^{(l)}_{i,j}\in\mathbb{R}^{d}$, where $l$ indexes the encoder layer, $i$ the utterance, $j$ the word, and $d$ the dimensionality of the layer's output (which is the same across all layers).
The duration values for each word, denoted $d_{i,j}=t^\text{end}_{i,j}-t^\text{start}_{i,j}$, are also retained for further analysis.

\subsubsection{Neutral--Emphasized Contrastive Pairing}

To assess the emphasis sensitivity of representations while controlling for contextual and speaker-dependent factors, a contrastive pairing set is constructed in which each pair comprises one emphasized and one neutral word representation. 
Pairs are sampled from the dataset such that speaker, word, and transcript identity are matched, differing only in emphasis label.
This yields 3,732 aligned neutral--emphasized pairings.

\subsection{Representation Analysis}

\subsubsection{Sample-Wise Cosine Similarity}
\label{sec:sample-wise-groups}
To evaluate how emphasis affects word representations, the distribution of cosine similarities between samples is analyzed. 
For each pair of aligned emphasized and neutral words, the cosine similarity between their representations is computed.
The neutral--emphasized pairwise similarities are compared with neutral--neutral similarity baselines.
If the emphasized and neutral variants of the same word are nearly identical ($\mathrm{cos}(\theta)\approx1$), this suggests that emphasis has little effect. 
If they are consistently less similar, this may indicate a systematic prosodic shift.  
Analyzing the distribution of these similarities across the dataset provides an interpretable measure of the sensitivity of the model's representations to emphasis. 
The means of these distributions are reported as summary metrics, denoted by $\theta_{AA}$ (neutral--neutral), $\theta_{BB}$ (emphasized--emphasized), and $\theta_{AB}$ (neutral--emphasized).

In addition, the distribution of cosine similarities between all unique residual pairs, $\mathbf{R}=\mathbf{B}-\mathbf{A}$, is analyzed. 
Its mean, denoted $\theta_{RR}$, is equivalent to the metric defined in \cite{fournier_analogies_2020}:
\begin{equation}
    \theta_{RR}=\frac{1}{2N(N-1)}\sum_{i<j}\mathrm{cos}({\mathbf{r}}_i, \mathbf{r}_j)
\end{equation}
where $\mathbf{r}_i=\mathbf{b}_i-\mathbf{a}_i$ is the residual vector for the $i$-th pair. 
This metric captures the second-order structure of the residuals, quantifying whether emphasis transformations encoded by the model are directionally consistent across different word instances.
However, because $\theta_{RR}$ involves comparisons over all residuals, it averages over potentially diverse lexical and speaker identities and may include variation uncorrelated with emphasis. 

To reduce the impact of such variance, a first-order directional consistency metric, denoted $\theta_{\hat{R}}$, is used, defined as the cosine similarity between each residual $\mathbf{r}_i$ and the mean residual vector $\bar{\mathbf{r}}$:
\begin{equation}
    \theta_{\hat{R}}^{(i)} = \cos(\mathbf{r}_i, \bar{\mathbf{r}}), \quad \bar{\mathbf{r}} = \frac{1}{N} \sum_i \mathbf{r}_i
\end{equation}
This reflects how well each individual transformation aligns with the average emphasis direction. 

\subsubsection{Dimension-Wise Variance via PCA}
To complement the sample-wise analysis, the variance across representation dimensions is examined. 
For this, Principal Component Analysis (PCA) \cite{jolliffe_principal_2016} is applied to the following representation spaces:
\begin{itemize}
    \item Neutral word representations $\mathbf{A}\in \mathbb{R}^{N\times d}$
    \item emphasized word representations $\mathbf{B}\in \mathbb{R}^{N\times d}$
    \item Concatenated representations $\mathbf{C}= [\mathbf{A}~|~\mathbf{B}] \in \mathbb{R}^{N\times 2d}$
    \item Residual vectors $\mathbf{R}=\mathbf{B}-\mathbf{A}\in \mathbb{R}^{N\times d}$
\end{itemize}

With $\lambda_i$ denoting the eigenvalue corresponding to the \mbox{$i$-th} principal component (PC), the explained variance ratio is defined as:
\begin{equation}
    v_i=\frac{\lambda_i}{\sum_j^d\lambda_j}
\end{equation}
The effective dimensionality, $D_{95 \%}$, is defined as the number of PCs needed to explain at least 95\% of the total variance:
\begin{equation}
    D_{95 \%}=\min \left\{k: \sum_{i=1}^{k} {v}_{i} \geq 0.95\right\}
\end{equation}
A higher $D_{95 \%}$ in $\mathbf{C}$ than in either $\mathbf{A}$ or $\mathbf{B}$ suggests that emphasis introduces additional structured variation in representation space, potentially aligned with a prosodic axis.
Additionally, a low $D_{95 \%}$ in $\mathbf{R}$ implies that the transformation from neutral to emphasized representations lies in a low-dimensional subspace, indicating that emphasis is encoded consistently across samples (generalized) rather than as a unique variant of each sample (memorized).

\subsubsection{Midpoint Centering}
PCA typically involves mean-centering the data, which removes any global offset in the covariance estimate. 
However, when applied to residual vectors $\mathbf{r}_i = \mathbf{b}_i-\mathbf{a}_i$, mean-centering alters the interpretation of the resulting PCs.
Let $\bar{\mathbf{r}} = \frac{1}{N}\sum_i \mathbf{r}_i$ denote the mean residual vector.
The centered residual is then:
\begin{equation}
 \label{eqn:collapse}
        \tilde{\mathbf{r}}_i = \mathbf{r}_i-\bar{\mathbf{r}}
                             = (\mathbf{b}_i-\bar{\mathbf{b}})-(\mathbf{a}_i-\bar{\mathbf{a}})
\end{equation}
This effectively centers each group ($\mathbf{A}$ and $\mathbf{B}$) independently, eliminating the global offset between the emphasized and neutral representations.
As a result, centering would remove the very structure under investigation.
Hence, the sets $\mathbf{A}$ and $\mathbf{B}$ are \textit{midpoint}-centered prior to analysis. 
For each sample,
\begin{equation}
    \hat{\mathbf{a}}_i = \mathbf{a}_i - \mathbf{m}, \quad \hat{\mathbf{b}}_i = \mathbf{b}_i - \mathbf{m},
\end{equation}
where $\mathbf{m}=\frac{1}{2}(\bar{\mathbf{a}}+\bar{\mathbf{b}})$.

\subsubsection{Reconstructing Duration Change from Residual Geometry}
To test whether the residuals $\mathbf{r}_i=\mathbf{b}_i-\mathbf{a}_i$ encode interpretable prosodic transformations, a regression task is used to reconstruct relative word-level duration change, as it is a known acoustic correlate and proxy of emphasis \cite{ladd_prosodic_2023}.
The relative duration change between emphasized and neutral instances of the same word is defined as:
\begin{equation}
    \delta_i = \frac{d_i^{\text{emph}}-d_i^{\text{neut}}}{d_{i}^{\text{neut}}}
\end{equation}
where $d^{\text{neut}}_i$ and $d^{\text{emph}}_i$ are word durations obtained from forced alignment (see Section~\ref{sec:representation-extraction}).
This ratio reflects how much longer the emphasized word is relative to the neutral baseline.  
A ridge regression model is then fit to predict $\delta_i$ from the \mbox{top-$k$} PCs of the residuals $\mathbf{r}_i$, and $R^2$ scores are reported.

The same regression task is repeated on the remaining representation spaces ($\mathbf{A},\mathbf{B},\mathbf{C}$).
Fitting on concatenated representations is expected to perform at least as well as $\mathbf{A}$ and $\mathbf{B}$, since the regressor has access to full information about both domains.
Higher predictive performance from the residual space would support the hypothesis that emphasis is encoded as a structured, low-dimensional transformation, potentially making non-linear perceptual effects in speech linearly accessible.

\subsubsection{Word Identity Prediction} 
To assess whether lexical information is accessible in representations, a simple word identity prediction task is performed. 
Similar to above, a logistic regression probing model is trained to predict word identity from representations. 

Applied to residual representations, this provides an approximate measure of disentanglement: if residuals capture only emphasis transformations, they should contain little to no information about the underlying word.  
Recent work has shown that lexical or paralinguistic features can be explicitly removed from speech representations via linear projection, yielding disentangled representations \cite{zhang_representation_2024}.
In contrast, the current experiment evaluates inherent disentanglement without additional fine-tuning of the residuals.

The logistic regression model is trained using standard cross-entropy loss and a fixed learning rate of $1 \times 10^{-4}$.  
The dataset contains 546 unique word classes.
Training is performed on 80\% of the pairs (2985), with accuracy evaluated on a 20\% held-out test set (747). 
This simple probe setup ensures that results reflect the information content of the representations rather than the capacity of the classifier.

The effective dimensionality required to achieve 95\% of the task-specific performance is also computed.
This is done by incrementally including the top-$k$ PCs and identifying the smallest $k$ for which cumulative performance reaches 95\% of the maximum, providing insight into how concentrated the investigated information is within each representation space.
This is then summarized using the area under the curve (AUC) over increasing $k$.

\subsubsection{Layer- and Model-Wise Comparison}
To investigate how emphasis sensitivity develops across layer-depth and training objectives, the duration change reconstruction and word identity prediction analyses are repeated across the following:
\begin{enumerate}
\item All encoder layers of each model,
\item A selection of pre-trained S3L models (e.g., wav2vec 2.0, HuBERT),
\item Fine-tuned variants, including:
\quad (a) ASR models, and 
\quad (b) a model fine-tuned for emphasis classification.
\end{enumerate}

This setup enables comparison of how task-relevant information is distributed across layers and whether fine-tuning shifts the encoding of emphasis from an implicit, low-dimensional transformation toward a more categorical or disentangled structure.

\section{Experiments}
\label{sec:experiments}

The analysis is demonstrated on layer 7 of wav2vec 2.0 as a worked example. 
It is then extended to all layers.
Finally, different models and fine-tuning objectives are compared.

\begin{table}[h!]
    \centering
    \caption{Summary of encoding properties for each group at a single layer. 
    Effective dimensionality $D_{95\%}$ is computed from the explained variance.
    Top-$k$ ($k=20$) correlation is the average absolute correlation with duration change.
    $R^2_\text{AUC}$ and $R^2_{95\%}$ is computed from regression onto duration change.
    WID$_\text{AUC}$ and WID$_{95\%}$ show the performance and effective dimension of the word reconstruction task}
    \begin{tabular}{l|ccccll}
        \textbf{Space} & \textbf{$D_{95\%}$} & Corr & \textbf{$R^2_{\mathrm{AUC}}$} & \textbf{$R^2_{95\%}$}  & WID$_\text{AUC}$ &WID$_{95\%}$\\
        \hline
        $\mathbf{A}$   & 308 & 0.31 & 0.60 & 382 & 0.66 &398\\
        $\mathbf{B}$   & 273 & 0.34 & 0.56 & 375 & 0.65 &417\\
        $\mathbf{C}$   & 473 & 0.33 & 0.66 & 370 & 0.66 &406\\
        $\mathbf{R}$   & 402 & 0.36 & \underline{0.71} & 341  & \underline{0.26} & 476\\
    \end{tabular}

    \label{tab:summary-w2v2-L7}
\end{table}

\subsection{Cosine Similarity Distributions}

The first experiment quantifies the extent to which emphasis alters word-level representations.
\begin{figure}[h!]
    \centering
    \includegraphics[width=0.75\linewidth]{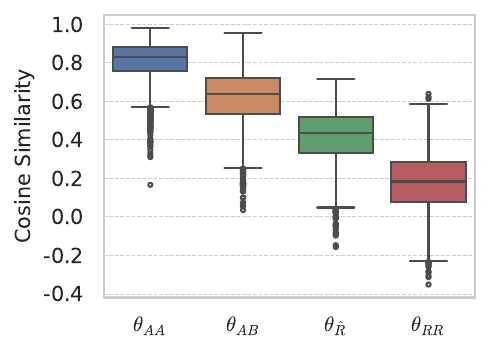}
    \caption{Cosine similarity distributions across neutral and emphasized word representations on wav2vec 2.0, layer 7.\\
    $\theta_{AA}$: neutral--neutral word pairs; $\theta_{AB}$: neutral--emphasized word pairs; $\theta_{\hat{R}}$: residuals aligned to the mean residual vector; $\theta_{RR}$: pairwise cosine similarity between residuals.}
    \label{fig:cosine-dists-w2v2}
\end{figure}
Figure~\ref{fig:cosine-dists-w2v2} suggests that emphasis induces a subtle but structured representational shift, with residual representations exhibiting both directional alignment and spread.

\subsection{Dimensionality Analysis}

\begin{figure}[h!]
    \centering
    \includegraphics[width=0.8\linewidth]{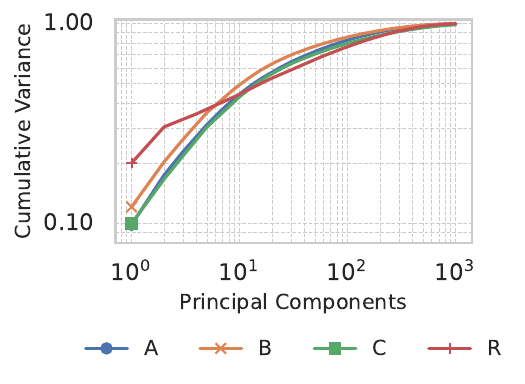}
    \caption{Cumulative variance explained over PCs.}
    \label{fig:pca-evrs-midpoint-w2v2}
\end{figure}

Figure~\ref{fig:pca-evrs-midpoint-w2v2} shows the cumulative variance explained by PCs in each representation space.
The residual space $\mathbf{R}$ appears more structured than the others in the early PCs, suggesting consistent variance between $\mathbf{B}$ and $\mathbf{A}$ along a low-dimensional subspace.

\subsubsection{Correlations Between PCs and Duration}

\begin{figure}[h!]
    \centering
    \includegraphics[width=0.7\linewidth]{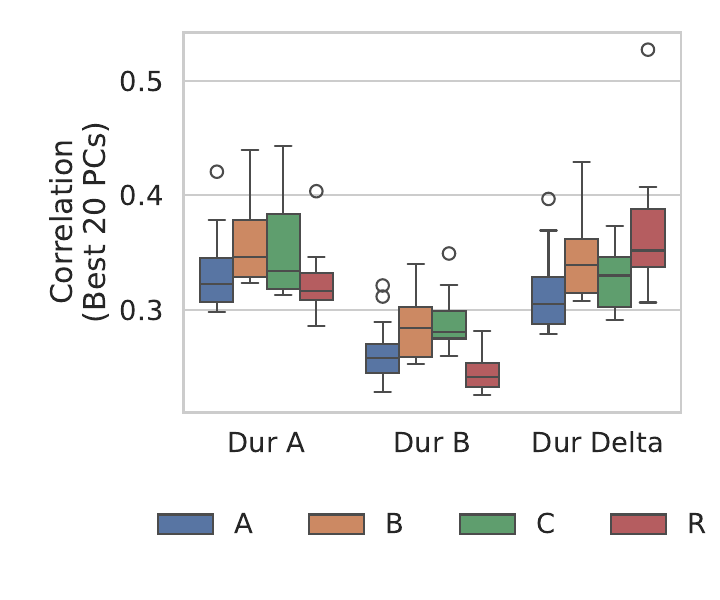}
    \caption{Top-20 ranked PCs correlated with neutral duration (Dur A), emphasized duration (Dur B), and percentage duration change (Dur $\delta$).
    }
    \label{fig:dur_corr_hist-w2v2}
\end{figure}
Figure~\ref{fig:dur_corr_hist-w2v2} suggests a stronger correlation with duration change, providing evidence that these PCs best explain the emphasis transformation. 
\begin{figure}[h!]
    \centering
    \includegraphics[width=0.65\linewidth]{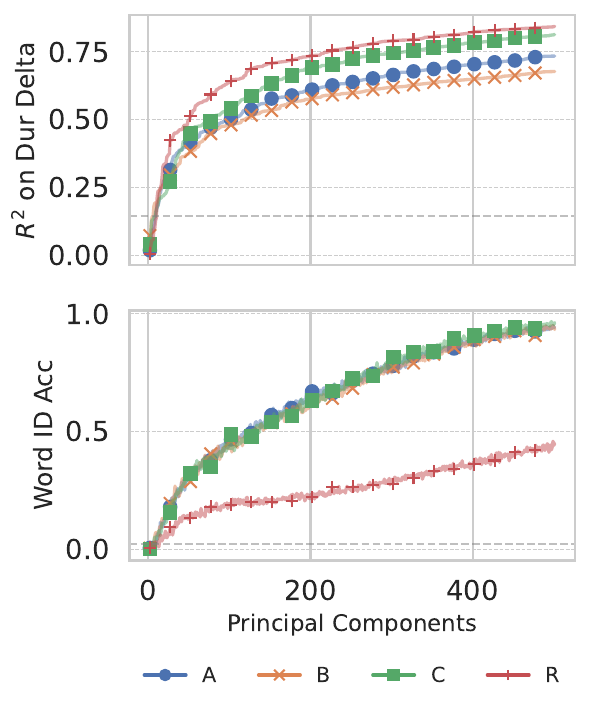}
    \caption{The residual vectors contain enough information to reconstruct duration change ($\delta$) but fail to predict word identity. 
    Conversely, duration change is less well recovered by the emphasized, neutral, or concatenated representations.}
    \label{fig:dur_r2-w2v2}
\end{figure}
Figure~\ref{fig:dur_r2-w2v2} illustrates the cumulative performance over PCs for both duration change reconstruction and word identity prediction.
These results, summarized in Table~\ref{tab:summary-w2v2-L7} by taking the AUC over PCs, show that the residual vectors retain enough information to recover duration change but not word identity, which remains accessible to the original representation spaces.

\subsection{Layer-Wise Comparison}

\begin{figure}[htbp!]
    \centering
    \begin{subfigure}{\linewidth}
        \centering
        \includegraphics[width=0.85\linewidth]{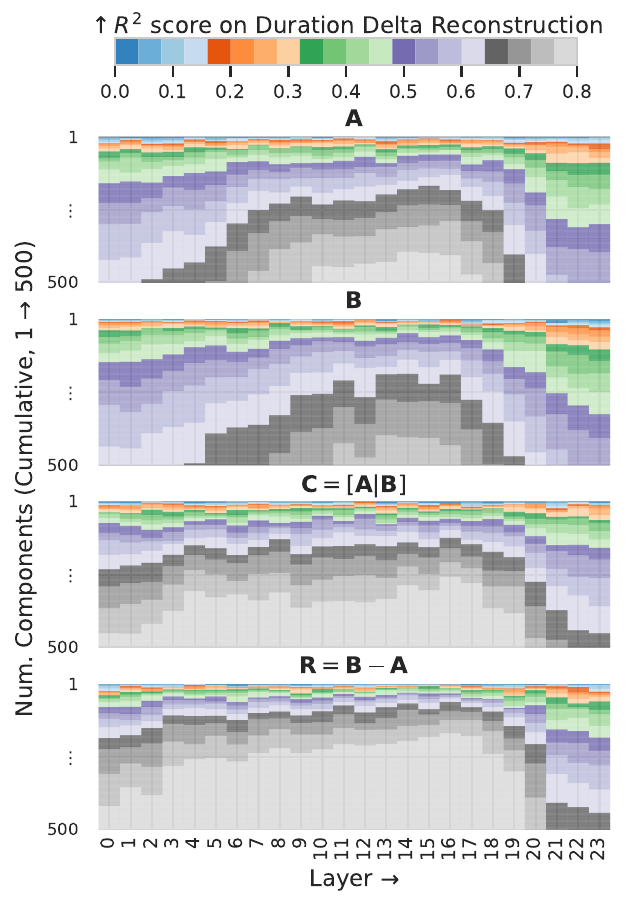}
        \caption{\(R^2\) score of delta duration reconstruction across layers and PCs.}
        \label{fig:durr_r2-layerwise}
    \end{subfigure}
    
    \vspace{0.5em}  

    \begin{subfigure}{\linewidth}
        \centering
        \includegraphics[width=0.85\linewidth]{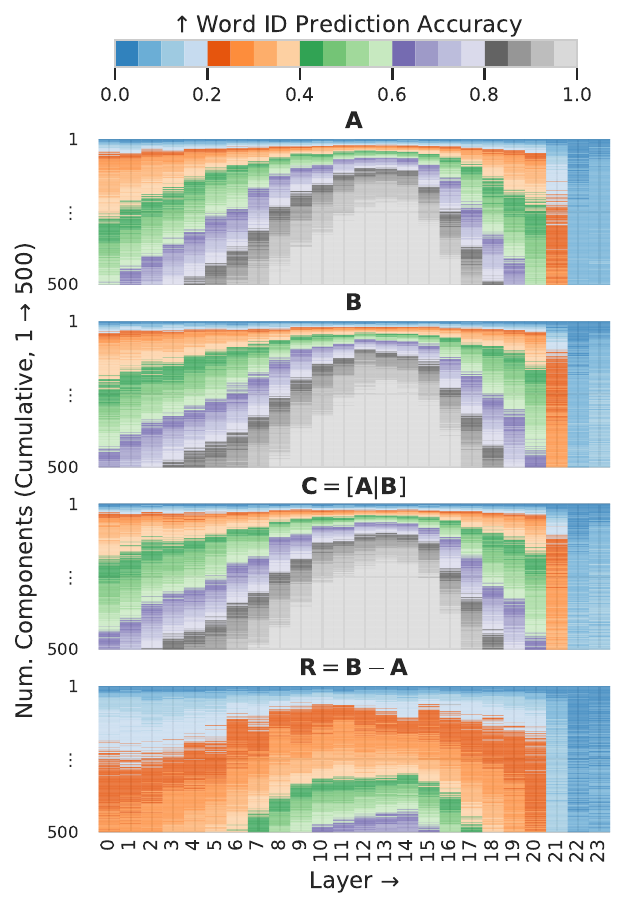}
        \caption{Word identity prediction accuracy across layers and PCs.}
        \label{fig:word_acc-layerwise}
    \end{subfigure}
    
    \caption{Layer-wise reconstruction and identity analysis. Top: regression-based reconstruction of emphasis variation. Bottom: word identity prediction performance.}
    \label{fig:layerwise_combined}
\end{figure}

\begin{table*}[htbp!]
    \centering
    \caption{Performance summary across models, showing area under the curve (AUC), effective dimensionality (Dim), and the corresponding layer of best performance for duration change ($\delta$) reconstruction and word identity prediction}
\begin{tabular}{l
  c@{ $\mid$ }c@{ $\mid$ }c  c@{ $\mid$ }c@{ $\mid$ }c  
  c@{ $\mid$ }c@{ $\mid$ }c  c@{ $\mid$ }c@{ $\mid$ }c  
}
\textbf{Model} &
\multicolumn{6}{c}{\textbf{Residuals ($\mathbf{R}$)}} &
\multicolumn{6}{c}{\textbf{Concatenated ($\mathbf{C}$)}} \\
& \multicolumn{3}{c}{Duration $\delta$} & \multicolumn{3}{c}{Word ID}
& \multicolumn{3}{c}{Duration $\delta$} & \multicolumn{3}{c}{Word ID} \\
& AUC & Dim & Layer & AUC & Dim & Layer & AUC & Dim & Layer & AUC & Dim & Layer \\
\hline
wav2vec 2.0\textsuperscript{\cite{baevski_wav2vec2_2020}}         & 0.75 & 298 & 16 & 0.36 & 454 & 13 & 0.69 & 325 & 16 & 0.86 & 178 & 13 \\
wav2vec 2.0 (ASR)   & 0.80 & 248 & 20 & 0.44 & 465 & 14 & 0.76 & 301 & 20 & 0.91 &  92 & 14 \\
XLS-R\textsuperscript{\cite{babu_xlsr_2022}}             & 0.76 & 309 & 32 & 0.26 & 476 & 32 & 0.70 & 341 & 35 & 0.84 & 206 & 28 \\
XLS-R (ASR)       & 0.82 & 277 & 36 & 0.74 & 244 & 30 & 0.78 & 248 & 47 & 0.95 &  48 & 31 \\
XLS-R (EC)\textsuperscript{\cite{de_seyssel_emphassess_2024}}
                 & 0.76 & 293 & 23 & 0.74 & 207 & 23 & 0.69 & 358 & 23 & 0.93 &  63 & 21 \\
HuBERT\textsuperscript{\cite{hsu_hubert_2021}}           & 0.72 & 304 & 22 & 0.69 & 271 & 23 & 0.67 & 351 & 14 & 0.91 &  91 & 23 \\
HuBERT (ASR)     & 0.81 & 263 & 23 & 0.69 & 295 & 22 & 0.73 & 337 & 22 & 0.94 &  59 & 22 \\
data2Vec\textsuperscript{\cite{baevski_data2vec_2022}}
                 & 0.82 & 253 & 21 & 0.40 & 473 & 21 & 0.73 & 357 & 21 & 0.90 & 119 & 20 \\
data2Vec (ASR)   & 0.83 & 240 & 20 & 0.39 & 457 & 21 & 0.75 & 296 & 20 & 0.90 & 117 & 21 \\
WavLM-Base\textsuperscript{\cite{chen_wavlm_2022}}       & 0.76 & 294 & 10 & 0.36 & 452 & 11 & 0.70 & 341 & 10 & 0.84 & 194 &  8 \\
WavLM-Base (ASR) & 0.77 & 293 & 10 & 0.34 & 499 &  7 & 0.72 & 330 & 11 & 0.87 & 132 &  8 \\
\end{tabular}

    \label{tab:model_summary}
\end{table*}
Figure~\ref{fig:layerwise_combined} shows the cumulative $R^2$ and accuracy scores for duration change reconstruction (top) and word identity prediction (bottom) across layers and PCs. 
The residual representations $\mathbf{R}$ achieve the highest reconstruction of duration change using fewer PCs, indicating that emphasis manifests as a structured, low-dimensional shift. 
In contrast, the residual yields near-zero performance on word identity prediction, indicating that lexical content is effectively removed.
Meanwhile, the concatenated representations $\mathbf{C}$ matches $\mathbf{R}$'s reconstruction performance, suggesting that the regressor can infer duration change from the full representations; yet only the residuals encode it directly and compactly.

Table~\ref{tab:model_summary} summarizes performance across models, comparing residual representations and concatenated representations for both duration change reconstruction and word identity prediction. 
Each entry shows the layer of best observed performance (AUC) and the number of PCs required to reach 95\% of that performance (Dim).
Across all models, residual representations consistently yield higher duration change reconstruction performance with lower effective dimensionality, indicating that emphasis is encoded as a structured, low-dimensional transformation. 
This effect is especially pronounced in fine-tuned ASR models, where residuals outperform raw representations while requiring fewer components.
Conversely, word identity prediction accuracy is substantially lower for residuals than for concatenated representations, suggesting that lexical content is largely absent, or at least obfuscated, in the residual space. 
This supports the hypothesis that residual representations primarily isolate prosodic variation rather than word-specific features.

\section{Discussion}
\label{sec:discussion}

The results provide strong evidence that emphasis is encoded as a structured, low-dimensional transformation within the internal representation spaces of the investigated speech models.
Residual vectors between aligned emphasized and neutral word representations show strong directional consistency and occupy significantly fewer dimensions than the full embedding space. 
This supports the hypothesis that emphasis is not memorized in a word-specific manner but instead emerges as a reusable prosodic shift in representation space.
Fine-tuning for ASR amplifies this effect: residuals become more predictive of duration change and less entangled with word identity, suggesting that ASR objectives may reinforce the accessibility of prosodic information. 
Word identity prediction from residuals remains low across models, further indicating that emphasis-related transformations are largely orthogonal to lexical encoding.

Interestingly, in the model fine-tuned for emphasis classification (XLS-R EC), duration change reconstruction is no longer dominated by the residual space. 
Emphasized words outperform neutral words in word identity prediction, suggesting the model may allocate more capacity on encoding emphasized content, potentially due to their relative rarity.

These findings echo results in style transfer and NLP analogy tasks, where residuals encode structured, interpretable variation. 
Unlike prior work that fine-tunes representations for disentanglement \cite{zhang_representation_2024}, this study shows that emphasis sensitivity can emerge inherently, particularly in middle-to-deep layers after fine-tuning.

\subsection{Limitations}

This work focuses on carefully controlled, aligned word pairs—matched by speaker, word, and sentence—to isolate the effect of emphasis. 
As a result, it does not explore how emphasis sensitivity behaves under relaxed conditions, such as varying speaker identity or contextual usage. 
In addition, all experiments are conducted on a benchmark synthetic dataset, which offers control over emphasis placement but may not fully capture the variability of natural speech. 
Finally, the analysis is limited to word-level emphasis, even though prosodic emphasis can span larger discourse units \cite{klewitz_quote_1999}. 
Investigating these broader and more variable conditions is left for future work. 
Nonetheless, the present findings demonstrate clear structure and interpretability under idealized settings, providing a strong foundation for further study.

\section{Conclusion}
\label{sec:conclusion}

This work investigates whether modern speech models encode prosodic emphasis as a structured transformation in representation space. 
Using a novel residual analysis framework that combines parameter-free geometric metrics with lightweight probing tasks, this study shows that S3L models and ASR-tuned models exhibit clear emphasis sensitivity. 
Residual vectors are directionally aligned, low-dimensional, and predictive of changes in duration.

Fine-tuning for ASR enhances this effect, making emphasis encoding more consistent and less entangled with lexical identity. 
These findings suggest that emphasis is not only accessible but also implicitly structured in speech representations, offering implications for prosody-aware speech modeling, analysis, and control.

\bibliography{IEEEabrv,references}

\end{document}